\documentclass{raa}           
\usepackage{graphicx,times}
\usepackage{amssymb,amsmath}
\usepackage{longtable}

\usepackage[pagebackref=true]{hyperref}

\begin{document}

   \title{Long-term Photometric Study of the Pre-main Sequence Star V1180 Cas}

 \volnopage{ {\bf 20XX} Vol.\ {\bf X} No. {\bf XX}, 000--000}
   \setcounter{page}{1}

   \author{Asen Mutafov
   \inst{1}, Evgeni Semkov\inst{1}, Stoyanka Peneva\inst{1}, Sunay Ibryamov
      \inst{2}
   }

   \institute{Institute of Astronomy and National Astronomical Observatory, Bulgarian Academy of Sciences, BG-1784, Sofia, Bulgaria {\it amutafov@asto.bas.bg}\\
        \and
              Department of Physics and Astronomy, Faculty of Natural Sciences, University of Shumen,
              115, Universitetska Str., 9700 Shumen, Bulgaria\\
\vs \no
   {\small Received 20XX Month Day; accepted 20XX Month Day}
}

\abstract{In this paper results from the optical photometric observations of the pre-main-sequence star V1180 Cas are reported.
The star is a young variable associated with the dark cloud Lynds 1340, located at a distance of 600 pc from the Sun in the star forming region in Cassiopeia.
V1180 Cas shows a large amplitude variability interpreted as a combination of accretion-induced and extinction-driven effects. 
Our data from VRI CCD photometric observations of the star are collected from September 2011 to February 2022. 
During our monitoring, we recorded several brightness dips with large amplitudes of up to 5 mag. (I-band).
At the same time, increases in brightness over periods of several weeks have also been recorded.
In this paper, we compare the photometric data obtained for V1180 Cas with observations of other low-mass pre-main sequence objects.
\keywords{stars: pre-main sequence -- stars: variables: T Tauri, Herbig Ae/Be -- stars: individual: V1180 Cas 
}
}

   \authorrunning{A. S. Mutafov et al. }            
   \titlerunning{Long-term Photometric Study of the Pre-main Sequence Star V1180 Cas}  
   \maketitle

%
\section{Introduction}           
\label{sect:intro}

One of the main fundamental characteristics of the young stellar objects is their photometric variability in the optical and near infrared range. 
It manifests itself as temporary dips in the brightness (eclipses), transient increases in brightness (outbursts), periodic or non-periodic brightness changes for a short or long-time scales. 
Photometric variability with amplitudes of different magnitudes and periodicity can be observed in both types of pre-main sequence (PMS) stars - the widespread low-mass ($\it M$ $\leq$ $2M_\odot$) T Tauri stars and the more massive Herbig Ae/Be (HAEBE) stars (Herbst et al. \cite{herb94}; Herbst et al. \cite{herb07}). 
It is widely accepted (Bertout \cite{bert89}; Appenzeller \& Mundt \cite{app89}) that PMS stars can be divided into two subclasses based on the presence of accretion disc - classical T Tauri (CTT) stars surrounded by a massive accretion disk and weak line T Tauri (WTT) stars without indications of disk accretion. 
While the variability of WTT stars is most often explained by rotational modulation of cool spots on their surface, CTT stars are characterized by variability with large amplitudes.
It can be explained by magnetically channeled accretion from the circumstellar disk onto the stellar surface. This explanation is summarized in the review work of Herbst et al. (\cite{herb07}).

A significant part of HAEBE stars and some early type CTT stars manifest strong photometric variability with sudden quasi-Algol drops in brightness and amplitudes up to $2\fm5$ ($V$) (Natta et al. \cite{natt97}; van den Ancker et al. \cite{anck98}; Herbst \& Shevchenko \cite{herb99}). 
This group of PMS objects with masses similar to or greater than the solar mass are named UXors after their prototype UX Orionis. 
They show an increase in polarization and specific color variability (“blueing effect”) at the deep minimums in brightness. 
One of the most widespread explanations of its variability is a variable extinction from dust clumps or filaments passing through the line of sight to the star (Dullemond et al. \cite{dull03}; Grinin et al. \cite{grin91}). 
Normally, when the star is covered by clouds of dust located along the line of sight, it becomes redder.
But when the obscuration rises sufficiently the scattered part of the light in the total observed light become considerable and the star color gets bluer.

Kun et al. (\cite{kunm94}) first recognized V1180 Cas as a young variable object with a strong H$_\alpha$ emission, associated with the dark cloud Lynds 1340 in the star forming region in Cassiopeia (Lynds ~\cite{lynd62}). 
Lynds 1340 is a small dark cloud with area 0.001 square degrees and opacity class 5, located at a distance of 600 pc from the Sun.
The photometric observations performed by Kun et al. (\cite{kunm11}) during the period from October 1999 to February 2011 show variability with an amplitude of about 6 mag. (I $ _C $ band).
Variability with such a large amplitude is consistent with that of known eruptive young stellar objects.
But the observed characteristic timescales of the faint and bright phases differ from these of the eruptive PMS stars from FU Ori (FUor) and EX Lupi (EXor) type.

Kun et al. (\cite{kunm11}) noted that the color magnitude diagram (I$_C$ versus R$_C$ - I$_C$) shows reddening while a weakening of the brightness of the star occurs. 
They determined the spectral type of V1180 Cas as K7-M0; the luminosity as $L/L_{\sun}$ $\approx$ 0.07 with an effective temperature T$_e$$_f$$_f$ = 4060 K of the K7 spectral type; the equivalent widths of H$_\alpha$ range from 300 to 900 $\AA$ and the mass accretion rate as $>$1.6x10$^{-7}$M$_{\sun}$/yr$^{-1}$ (Ca $_I$$_I$ $\lambda$ 8542).
The authors suggest that it is necessary to refine the current picture of the mechanisms of episodic accretion bursts at the time. They presume that V1180 Cas may be a new member of an unclassified eruptive young stars.

Near-IR monitoring of the object for the period September-August 2013 was reported by Antoniucci et al. (\cite{anto13}). 
The authors recognized two major outburst events and plot the [J-H] vs. [H-K] two-color diagram. According to them the description of the color variation cannot be pure extinction alone, which is in agreement with the conclusions of Kun et al. (\cite{kunm11}).
Using their optical and near-IR emission observations Antoniucci et al. (~\cite{anto14}) calculated a mass accretion rate of 3 x10$^{-8}$M$_{\sun}$/yr$^{-1}$.
The first X-ray detections of V1180 Cas are reported in the paper of Antoniucci et al. (\cite{anto15}). 
The authors have performed simultaneous observations – X-ray, JHK photometry and J-band spectroscopy. 
From the observed Chandra signal the authors estimate the X-ray luminosity L$_X$(0.5-7 keV) in the range 0.8 $\div$ 2.2 x10$^{30}$ $\rm erg~s^{-1}$. 
Based on the spectral energy distribution the authors inferred a stellar luminosity of 0.8-0.9 L$_{\sun}$. Furthermore, the measured X-ray flux emission levels are in the order of  5 x10$^{30}$ $\div$ 1 x10$^{31}$ $\rm erg~s^{-1}$. The authors concluded that the photometric behavior of V1180 Cas might be explained by a combination of accretion-induced and extinction-driven effects.

In our first paper (Mutafov et al. ~\cite{muta18}) we reported results from VRI photometric monitoring of V1180 Cas in the period from September 2011 to April 2018.
During this period we observed drops in stellar brightness with amplitude up to 3 mag. (Ic).
In this paper, we present data from the continuation of this multicolor optical monitoring until February 2022 and an analysis of the observed photometric variability of the star.


\section{Observations}
\label{sect:Obs}

The CCD observations of V1180 Cas were performed in two observatories with four telescopes: the 2-m Ritchey-Chr\'{e}tien-Coud\'{e}, the 50/70-cm Schmidt and the 60-cm Cassegrain telescopes of the Rozhen National Astronomical Observatory (Bulgaria) and the 1.3-m Ritchey-Chr\'{e}tien telescope of the Skinakas Observatory of the University of Crete (Greece)\footnote{Skinakas Observatory is a collaborative project of the University of Crete, the Foundation for Research and Technology - Hellas, and the Max-Planck-Institut f\"{u}r Extraterrestrische Physik.}. 
The observations were performed with five different types of CCD cameras: ANDOR iKon-L and Vers Array 1300B at the 2-m RCC telescope, ANDOR DZ436-BV at the 1.3-m RC telescope, FLI PL16803 at the 50/70-cm Schmidt telescope, and FLI PL9000 at the 60-cm Cassegrain telescope. 
The technical parameters and specifications of the used CCDs are summarized in Table ~\ref{tab1}.
Observational procedure and data reduction process are described in Ibryamov et al. (~\cite{ibry15}). 
All frames were taken through a standard Johnson-Cousins set of filters.
The observations in all filters (VRI) are not simultaneous and have total duration of 20 - 30 minutes. During this time there is no significant change in the brightness of the star which could lead to significant changes of the color index.

   \begin{table*}
   \small
 \begin{center}
   \caption[]{CCD cameras and optical specifications\label{tab1}}
   \begin{tabular}{llllllllllll}

            \hline
            \noalign{\smallskip}

 Telescope  & CCD Camera Type& Size & Field  &Pixel size&Scale      & RON        &Gain \\
 &                &(px)      &(arcmin)&($\mu$m)   &(\arcsec/px)&($e^-$rms)&($e^-$/ADU)\\  
            \noalign{\smallskip}
            \hline
            \noalign{\smallskip}
2m RCC& Vers Array 1300B&1340$\times$1300& 5.8$\times$5.6&20.0&0.26&2.00&1.0\\
2m RCC& ANDOR iKon-L&2048$\times$2048& 6.0$\times$6.0&13.5&0.17&6.90&1.1\\
Schmidt& FLI PL16803&4096$\times$4096& 73.8$\times$73.8&9.0&1.08&9.00&1.0\\
60cm Cass& FLI PL9000&3056$\times$3056& 16.8$\times$16.8&12.0&0.33&8.50&1.0\\
1.3m RC&ANDOR DZ436-BV&2048$\times$2048& 9.6$\times$9.6&13.5&0.28&8.14&2.7\\

\noalign{\smallskip}
            \hline

         \end{tabular}
  \end{center}
   \end{table*}

\begin{table}
\bc
\begin{minipage}[]{100mm}
\caption[]{Coordinates and the photometric data for the VRI comparison sequence\label{tab2}}\end{minipage}
\setlength{\tabcolsep}{2.5pt}
\small
 \begin{tabular}{ccccccccccccc}
  \hline\noalign{\smallskip}
Standard&$\alpha (2000)$&$\delta (2000)$&V&$\sigma_V$&R&$\sigma_R$&I&$\sigma_I$\\

  \hline\noalign{\smallskip}

A&02 33 25.2&72 43 54.9&16.84&0.04&15.78&0.03&14.39&0.03\\
B&02 32 58.0&72 42 54.5&17.64&0.03&16.29&0.04&14.68&0.03\\
C&02 33 06.0&72 41 58.8&18.68&0.05&17.58&0.03&16.46&0.04\\
D&02 33 11.3&72 43 37.2&-&-&18.79&0.07&16.54&0.07\\
E&02 32 48.7&72 42 45.0&18.69&0.05&17.60&0.03&16.56&0.03\\
F&02 33 06.7&72 45 20.9&19.63&0.03&18.59&0.07&17.51&0.07\\
G&02 33 03.2&72 42 11.9&-&-&19.15&0.07&17.67&0.06\\
H&02 32 52.0&72 44 32.9&19.73&0.05&18.72&0.06&17.68&0.05\\

  \noalign{\smallskip}\hline
\end{tabular}
\ec
\end{table}

\begin{figure} 
   \centering
   \includegraphics[width=10.0cm, angle=0]{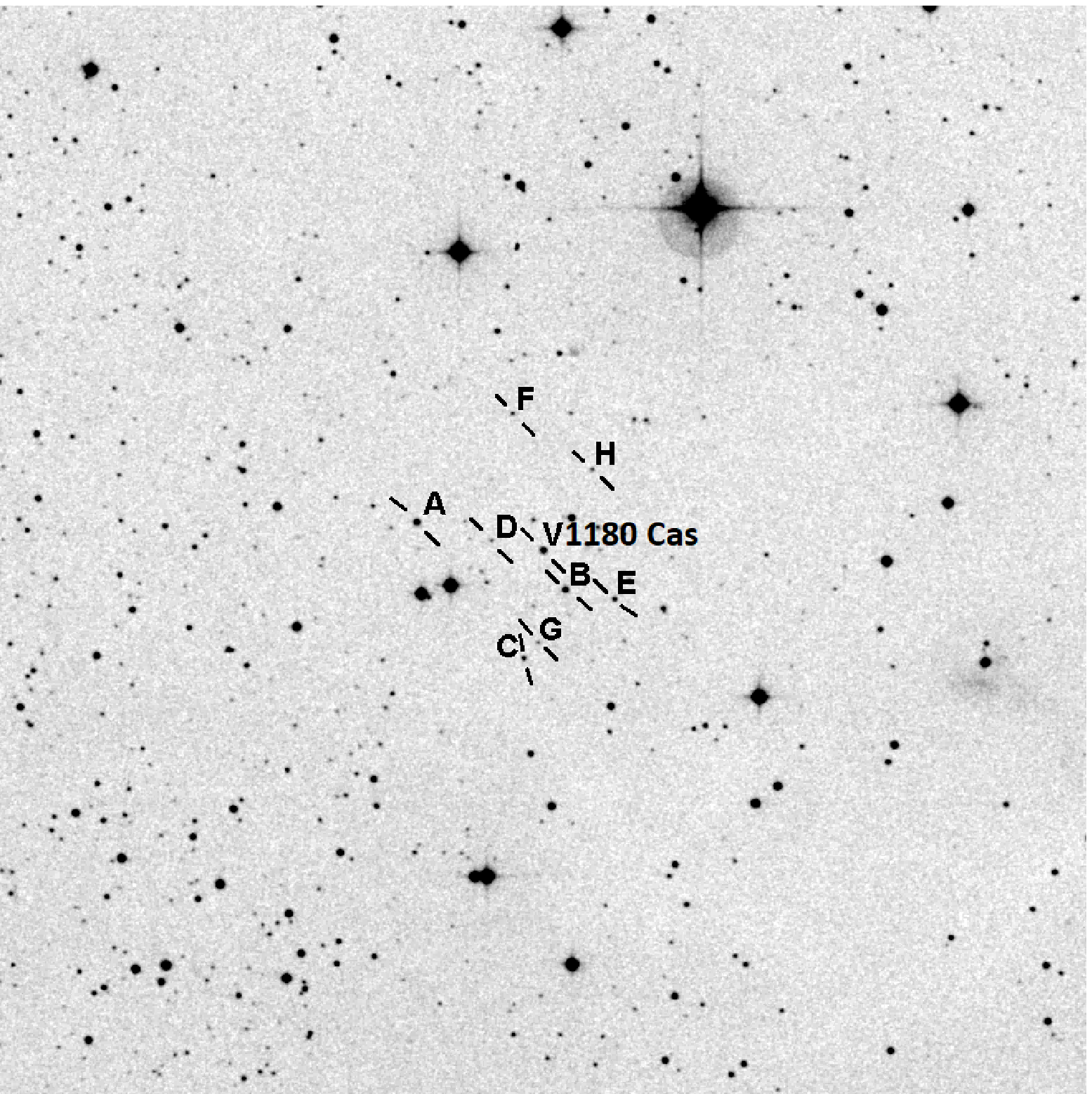}
   \caption{The finding chart of the comparison sequences. } 
   \label{Fig1}
   \end{figure}

In order to facilitate transformation from instrumental measurement to the standard Johnson-Cousins system eight stars in the field of V1180 Cas were calibrated in VRI bands. Calibration was made during six clear nights in 2011, 2012 and 2015 with the 1.3-m RC telescope of the Skinakas Observatory. Standard stars from Landolt (~\cite{land92}) were used as reference. The finding chart of the comparison sequences is presented in Fig. ~\ref{Fig1}. The comparison stars are labeled from A to H in order of their V magnitudes. The field is $15\arcmin\times15\arcmin$, centered around V1180 Cas. North is at the top and east to the left. The chart is retrieved from the STScI Digitized Sky Survey Second Generation Red.
Table ~\ref{tab2} contains the coordinates and the photometric data for the VRI comparison sequence. 
Our data was analyzed using fixed apertures for V1180 Cas which was chosen to be 4\arcsec radius. The background annulus was taken from 13\arcsec to 17\arcsec.

\section{Results and Discussion}
\label{sect:Discussion}

The results of our multicolor photometric CCD observations of V1180 Cas in the period from September 2011 to February 2022 are summarized in Table 3. 
The columns provide the Julian date (JD) of observation, VRI magnitudes and the telescope used. In the column Telescope, the abbreviation 2-m denotes the 2-m Ritchey-Chr\'{e}tien-Coud\'{e} telescope, Sch - the 50/70-cm Schmidt telescop, 60-cm - the 60-cm Cassegrain telescope and 1.3-m - the 1.3-m Ritchey-Chr\'{e}tien telescope. 
The values of instrumental errors are in the range $0\fm01$-$0\fm02$ (for I and R) and $0\fm01$-$0\fm03$ (for V). The VRI-light curves of V1180 Cas during the period of our observations are plotted in Fig. 2. 
In most cases the size of error bars are smaller than the size of the symbols used.

{\footnotesize
\begin{longtable}{cccccccccc} 


\caption{VRI photometric observations of V1180 Cas}\label{tab3}\\ 

\hline
\noalign{\smallskip}  
J.D. (24…)&V&R&I&Tel&J.D. (24…)&V&R&I&Tel\\ 
\noalign{\smallskip}  
\hline
\endfirsthead
\caption{VRI photometric observations of V1180 Cas. Continued.}\\
\hline
\noalign{\smallskip}  
J.D. (24…)&V&R&I&Tel&J.D. (24…)&V&R&I&Tel\\ 
\noalign{\smallskip}  
\hline

\noalign{\smallskip}  
\endhead
\hline

\endfoot
\noalign{\smallskip}
55824.45&18.68&17.26&15.86&1.3-m&57603.49&17.84&16.50&15.21&2-m\\
55842.39&18.66&17.33&16.02&1.3-m&57605.49&18.07&16.57&15.25&Sch\\
55848.40&18.60&17.16&15.76&1.3-m&57607.47&17.85&16.45&15.14&Sch\\
55865.40&&17.42&16.06&2-m&57664.46&17.84&16.36&15.14&Sch\\
55866.50&18.58&17.19&15.85&2-m&57698.36&18.01&16.66&15.33&Sch\\
55892.34&&17.15&15.76&2-m&57714.43&17.81&16.59&15.31&2-m\\
55896.35&&16.88&15.58&Sch&57715.42&18.01&16.76&15.45&2-m\\
55925.33&&16.87&15.55&Sch&57716.43&17.93&16.73&15.48&2-m\\
56122.52&18.46&16.62&15.30&Sch&57756.33&17.87&16.45&15.20&Sch\\
56123.51&&16.98&15.60&Sch&57781.26&17.96&16.55&15.20&Sch\\
56141.54&&16.77&15.44&1.3-m&57782.33&17.85&16.60&15.23&2-m\\
56159.50&&17.20&15.86&Sch&57785.28&19.05&17.61&16.25&2-m\\
56160.50&&17.25&15.88&Sch&57786.31&18.80&17.32&15.87&2-m\\
56161.52&&17.15&15.79&Sch&57801.26&&16.60&15.29&Sch\\
56173.47&18.72&17.06&15.74&1.3-m&57817.23&&16.43&15.16&Sch\\
56174.48&18.90&17.15&15.77&1.3-m&57893.54&17.56&16.36&15.10&2-m\\
56182.41&18.40&16.87&15.54&1.3-m&57904.54&17.66&16.30&15.10&Sch\\
56183.46&17.86&16.51&15.23&1.3-m&57968.49&18.80&17.24&15.84&Sch\\
56193.45&18.01&16.59&15.28&1.3-m&57969.47&18.56&16.98&15.57&Sch\\
56193.47&18.04&16.60&15.27&Sch&58011.43&18.70&17.23&15.82&Sch\\
56194.46&&16.75&15.43&Sch&58012.43&19.40&17.91&16.47&Sch\\
56214.33&17.86&16.59&15.26&2-m&58013.46&-&18.52&17.31&Sch\\
56249.37&&16.98&15.65&Sch&58043.41&&17.85&16.43&Sch\\
56275.37&18.22&16.87&15.57&2-m&58080.39&19.65&18.23&16.95&Sch\\
56328.37&&17.76&16.44&Sch&58081.40&20.24&18.58&17.21&Sch\\
56329.29&&17.65&16.30&Sch&58109.42&&18.44&17.80&Sch\\
56356.29&&17.62&16.14&60-cm&58113.34&20.50&18.72&17.81&Sch\\
56357.29&&17.53&16.10&60-cm&58114.35&&18.77&18.00&Sch\\
56371.29&18.49&17.13&15.85&2-m&58220.58&18.14&16.75&15.34&Sch\\
56428.52&&17.36&16.11&60-cm&58220.59&18.37&16.80&15.39&2-m\\
56430.50&&&15.98&60-cm&58278.52&18.00&16.73&15.47&Sch\\
56432.49&&17.32&15.83&60-cm&58343.47&18.00&16.60&15.24&Sch\\
56444.55&&18.02&16.41&Sch&58344.47&17.97&16.62&15.26&Sch\\
56478.53&18.85&17.36&16.00&2-m&58346.49&18.05&16.75&15.34&2-m\\
56507.52&19.76&18.32&17.30&2-m&58363.50&17.88&16.54&15.16&Sch\\
56508.51&19.53&18.01&16.95&2-m&58364.44&&16.74&15.38&Sch\\
56509.47&&17.71&16.59&Sch&58365.45&18.30&16.99&15.63&2-m\\
56510.49&&17.74&16.71&Sch&58428.32&&17.90&17.08&Sch\\
56510.55&&17.68&16.64&60-cm&58492.26&&18.41&17.75&Sch\\
56511.51&&&16.99&Sch&58496.28&&&18.01&Sch\\
56512.49&&17.94&17.18&Sch&58543.38&&&19.65&Sch\\
56540.46&&19.17&18.66&Sch&58544.24&&&19.58&2-m\\
56541.49&&19.23&18.73&Sch&58604.56&&&17.20&Sch\\
56543.48&20.82&19.62&&2-m&58606.56&19.32&18.11&16.86&2-m\\
56544.45&20.81&19.33&18.20&2-m&58666.52&&17.34&15.92&Sch\\
56553.45&19.63&17.87&16.44&1.3-m&58667.55&&17.48&15.92&Sch\\
56577.54&&16.89&15.67&60-cm&58690.52&18.08&16.73&15.32&2-m\\
56578.55&&17.02&15.68&60-cm&58691.56&17.96&16.68&15.25&2-m\\
56604.48&&16.98&15.57&60-cm&58704.50&18.90&17.39&15.84&Sch\\
56655.32&&17.98&16.93&Sch&58705.49&18.60&17.27&15.72&Sch\\
56656.47&&17.95&16.88&Sch&58706.50&18.80&17.34&15.70&Sch\\
56657.27&&18.05&17.00&Sch&58707.52&18.80&17.29&15.71&Sch\\
56681.32&&16.83&15.55&Sch&58726.48&18.80&17.40&15.95&2-m\\
56694.26&17.93&16.63&15.40&2-m&58727.51&&17.54&16.03&2-m\\
56715.40&17.83&16.42&15.18&Sch&58728.59&&17.82&16.30&2-m\\
56738.36&&16.57&15.28&Sch&58729.48&19.34&17.76&16.23&2-m\\
56899.46&18.24&16.69&15.40&1.3-m&58730.51&19.10&17.58&16.09&2-m\\
56988.29&&17.07&15.66&Sch&58730.57&&17.65&16.13&Sch\\
57005.34&&17.00&15.58&Sch&58758.26&&18.52&17.29&Sch\\
57006.43&&16.71&15.39&Sch&58759.41&&18.27&17.06&Sch\\
57016.39&19.95&18.21&16.78&2-m&58864.32&&&18.95&Sch\\
57074.23&&&15.83&Sch&58865.33&&&18.35&Sch\\
57164.55&&17.88&16.21&Sch&58867.29&20.75&19.62&18.90&2-m\\
57187.55&18.65&16.97&15.55&2-m&58870.32&&19.30&19.00&Sch\\
57190.54&18.44&17.01&15.55&2-m&58909.25&&19.37&18.20&Sch\\
57220.52&&17.05&15.67&Sch&58993.55&&&18.22&Sch\\
57221.54&&17.10&15.67&Sch&58993.53&&19.32&18.27&2-m\\
57223.54&18.36&16.92&15.56&2-m&59041.51&19.10&17.46&16.11&Sch\\
57224.53&18.27&16.84&15.45&2-m&59042.50&19.37&17.91&16.79&Sch\\
57246.53&18.88&17.41&15.84&1.3-m&59059.44&&18.39&17.24&Sch\\
57247.48&18.64&17.19&15.65&1.3-m&59060.51&&18.05&16.92&Sch\\
57259.45&&16.84&15.33&Sch&59075.34&18.83&17.46&16.14&2-m\\
57260.47&&16.92&15.48&Sch&59101.50&&15.55&14.29&Sch\\
57271.49&18.64&17.24&15.79&2-m&59102.40&16.43&15.52&14.30&2-m\\
57272.47&18.25&17.04&15.59&2-m&59103.43&17.50&16.38&15.08&2-m\\
57330.34&17.93&16.55&15.24&Sch&59104.45&&16.80&15.51&Sch\\
57331.36&17.89&16.60&15.23&Sch&59105.47&17.22&16.13&14.83&Sch\\
57332.35&18.03&16.65&15.28&Sch&59109.44&17.75&16.55&15.37&Sch\\
57333.33&17.93&16.52&15.22&Sch&59176.34&18.65&17.22&15.60&Sch\\
57334.32&17.87&16.53&15.20&Sch&59177.36&18.58&17.06&15.48&Sch\\
57369.31&18.42&17.13&15.49&2-m&59220.31&20.10&18.68&17.21&2-m\\
57370.37&19.08&17.60&16.24&2-m&59250.32&18.88&17.34&15.86&2-m\\
57371.34&18.65&17.22&15.79&2-m&59251.27&19.02&17.48&16.02&2-m\\
57372.33&19.20&17.57&16.20&Sch&59314.58&18.16&16.81&15.45&2-m\\
57374.34&18.68&17.17&15.77&Sch&59402.53&17.97&16.76&15.41&2-m\\
57390.23&18.24&16.76&15.43&Sch&59403.51&17.33&16.15&14.90&Sch\\
57425.39&17.92&16.45&15.21&Sch&59426.50&16.91&15.95&14.69&2-m\\
57426.40&17.94&16.46&15.21&Sch&59428.52&17.32&16.30&14.99&2-m\\
57427.37&17.86&16.35&15.16&Sch&59441.50&16.97&15.83&14.57&Sch\\
57483.60&17.75&16.36&15.03&2-m&59468.45&&16.23&14.93&Sch\\
57484.60&17.81&16.41&15.10&2-m&59523.33&&16.90&15.58&Sch\\
57522.54&17.93&16.37&15.22&Sch&59549.37&17.37&16.15&14.91&Sch\\
57523.54&18.22&16.50&15.32&Sch&59550.34&17.36&16.16&14.86&Sch\\
57565.53&&16.12&14.95&Sch&59582.32&17.42&16.26&14.99&Sch\\
57581.52&18.03&16.42&15.17&Sch&59605.30&18.24&17.10&15.80&2-m\\
57582.55&17.89&16.46&15.16&Sch&59636.25&&17.55&16.60&Sch\\
57583.53&17.72&16.35&15.09&Sch&\\

\hline
\end{longtable}}

\begin{figure} 
   \centering
   \includegraphics[width=12.0cm, angle=0]{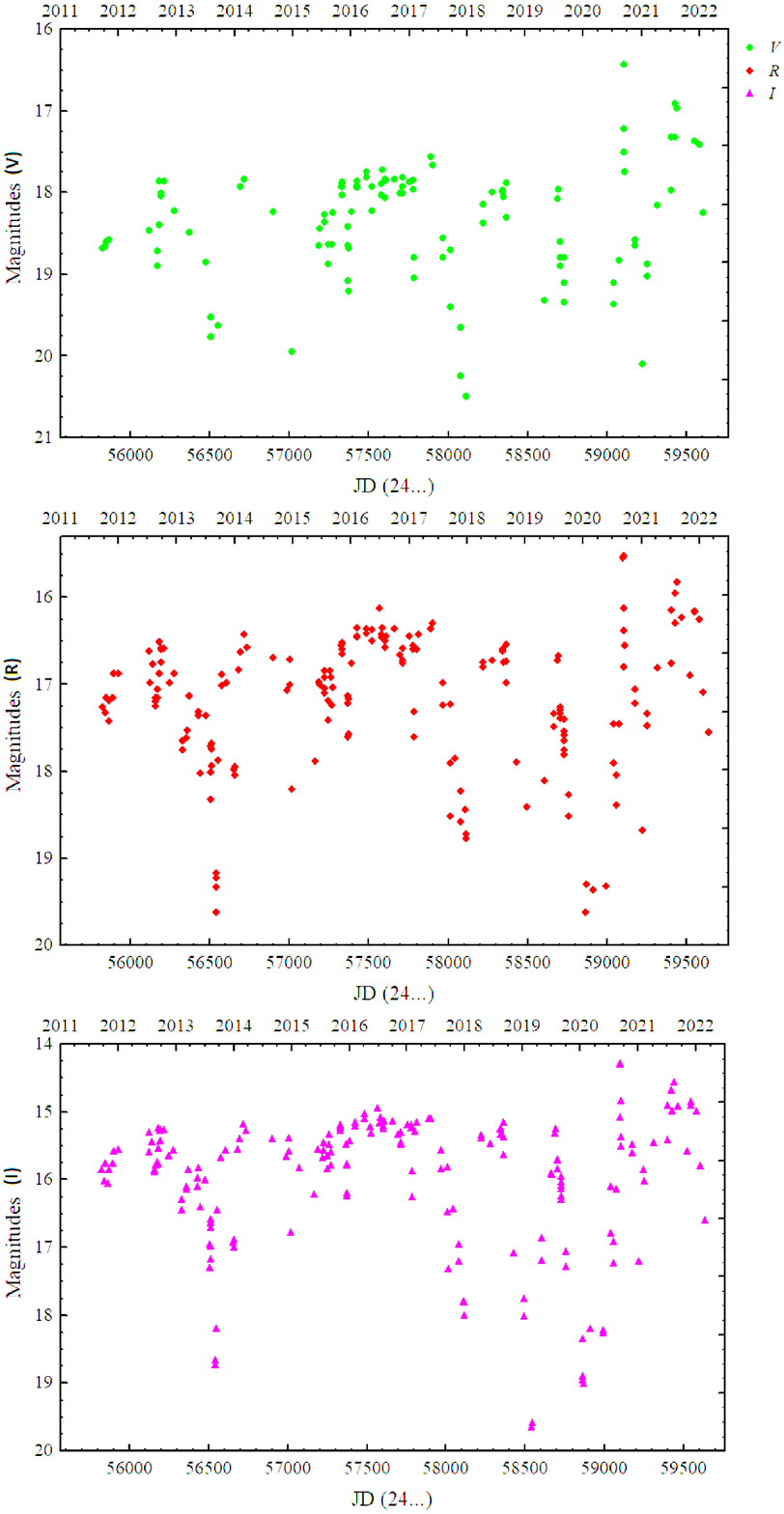}
   \caption{VRI light curves of V1180 Cas in the period September 2011 - February 2022. } 
   \label{Fig2}
   \end{figure}

During the period from 2011 to 2020 our data shows very strong photometric variability with a large amplitude variations ($\Delta$ I $\sim$ 5 mag). 
The same brightness variability was also registered in the previous studies of Kun et al. (\cite{kunm11}), Antoniucci et al. (\cite{anto13}, \cite{anto14}) and Lorenzetti et al. (\cite{lore15}).
Most of the time, the I-band brightness of the star is kept in the range of 15-16 mag., which is considered to be the maximum light in the previous studies.
However, during these periods there are changes in brightness with small amplitudes, which is characteristic of T Tau stars.

Our photometric data over the same time period also show another type of variability, deep declines in brightness for which no periodicity is observed.
We have registered four deep minimums in brightness in the light curve of V1180 Cas: the first deep minimum is registered on September 2013, the second on December 2017, the third on February-March 2019 and fourth on January 2020. 
Therefore, the long-term light curve of V1180 Cas is similar to that of other low-mass young stars as GM Cep (Semkov and Peneva ~\cite{semk12}; Semkov et al. ~\cite{semk15}; Mutafov et al. ~\cite{muta20}), 2MASS J22534654+6234582 (Ibryamov et al. ~\cite{ibry20}) and FHO 27 (Findeisen et al. ~\cite{find13}, Ibryamov et al. ~\cite{ibry19}). 
All of these objects show typical UXor variability, but are T Tauri stars from late spectral types. 

The color-magnitude diagrams of V1180 Cas V versus (V-R), V versus (V-I) and R versus (R-I) are shown in Fig. 3. 
The collected multicolor photometric data shows the typical for UXors color reversal during the minimums in brightness. 
This is a manifestation of the so-called “blueing” effect, the color of the star gets bluer in the minima of its brightness, in accordance with the model of dust-clump obscuration. 
The figure shows that, for each of the color diagrams a point of color reversal is observed at different star brightnesses: in the V to (V-R) diagram, the point of reversal is observed at V about 19.0 mag, in the V to (V-I) diagram at V about 19.2 mag and in the R to (R-I) diagram at R about 17.5 mag. In the color-magnitude diagrams of V1180 Cas like in the case of GM Cep and V1184 Tau the "blueing" effect is observed, besides in V-R color, in the R-I color, too (Semkov et al. ~\cite{semk13}, Semkov et al. ~\cite{semk15}, Mutafov et al. ~\cite{muta20}, Mutafov et al. ~\cite{muta22}). 

Since the fall of 2020, there has been a significant change in the photometric behavior of V1180 Cas.
We registered two increases in brightness (local maximums of brightness): the first one on September 2020 and the second on July/August 2021.
In this case, the increase in brightness seems to be caused by increased accretion.
Evidence of this is the decrease in the color indices (V-R and V-I) of the star (Fig. 3), during the increase in brightness.

PMS stars are characterized by different types of variability.
In many cases, two or more different types of variability can be observed for the same star.
We know objects that show both an increase in brightness due to increased accretion and a decrease in brightness caused the variable extinction such as V582 Aur (Semkov et al. ~\cite{semk13}, \'{A}brah\'{a}m et al. ~\cite{abra18}).
The assumption that the observed variations in the brightness of V1180 Cas are a combination of variable accretion and variable extinction in the line of sight was made by Kun et al. (\cite{kunm11}) and Antoniucci et al. (\cite{anto15}).
But using data from multicolor photometry we can distinguish the two phenomena over different periods of time.

The UXor phenomenon has been discovered and defined in young stars with high and medium masses (HAEBE stars and early type CTT stars).
But recently, a large number of small-mass objects have been discovered that also show this type of variability (Semkov et al. ~\cite{semk13}, Ibryamov et al. ~\cite{ibry15}, Semkov et al. ~\cite{semk15}, Mutafov et al. ~\cite{muta22}).
This phenomenon is especially common among stars from young stellar clusters and stellar associations (Findeisen et al. ~\cite{find13}, Barsunova et al ~\cite{bars15}). 
This shows that the processes of formation of stars with different masses proceed similarly.
A significant part of the protostellar gas-dust cloud remains in the vicinity of the newly formed stars and causes variable extinction.

\begin{figure} 
   \centering
   \includegraphics[width=12.0cm, angle=0]{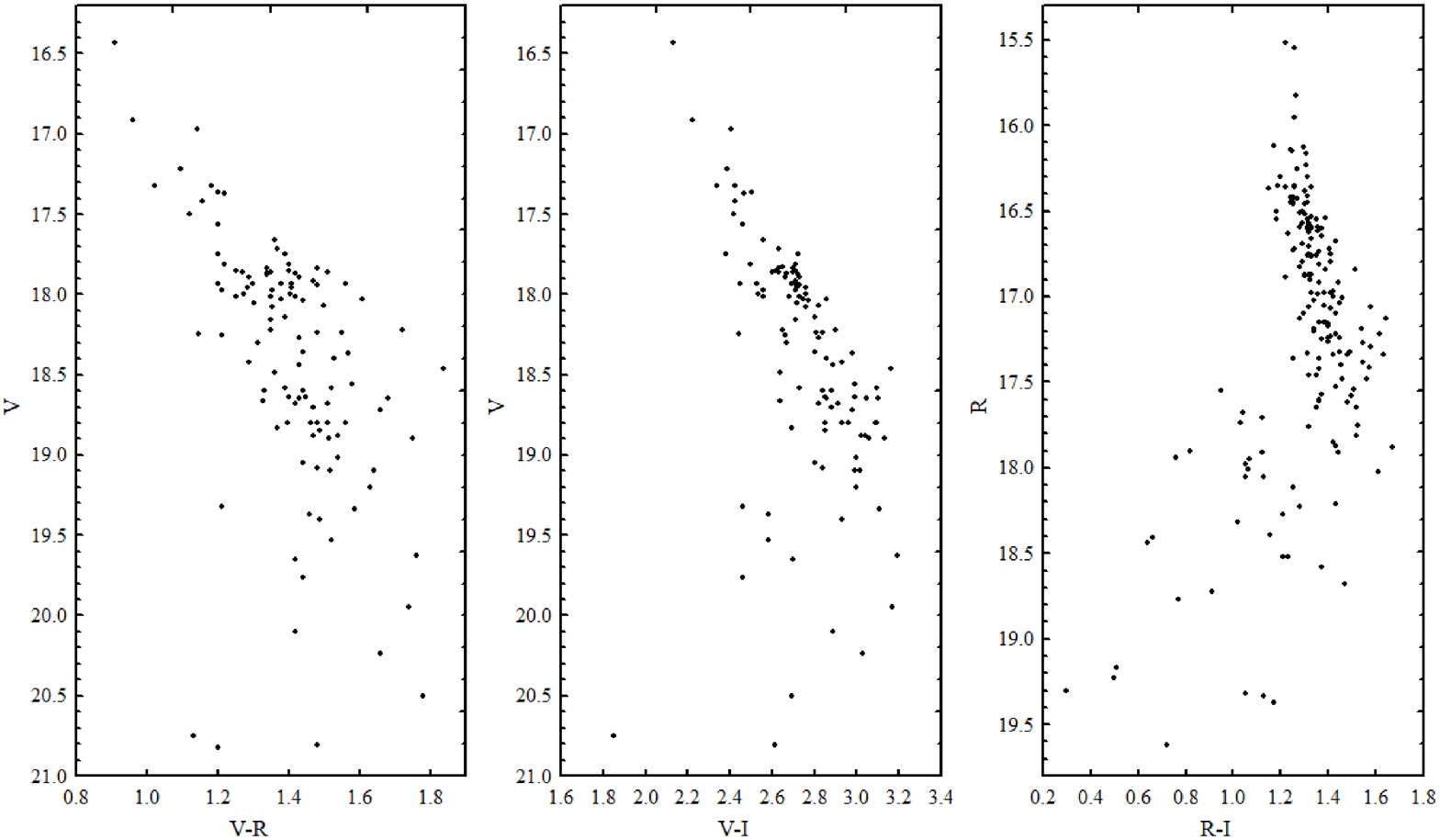}
   \caption{The color-magnitude diagrams of V1180 Cas in the period of observations September 2011 - February 2022. } 
   \label{Fig3}
   \end{figure}

\section{Conclusions}
\label{sect:Conclusions}

The last collected multicolored photometric data confirms that outside the deep minimums V1180 Cas shows significant variations of brightness lasting days and months. The same data, again, confirms that the variability of the star is dominated by the variable extinction. 
The VRI light curves of the V1180 Cas are similar to other objects - GM Cep (Semkov and Peneva ~\cite{semk12}; Semkov et al. ~\cite{semk15}; Mutafov et al. ~\cite{muta22}), V1184 Tau (Mutafov et al. ~\cite{muta20}), 2MASS J22534654+6234582 (Ibryamov et al. ~\cite{ibry20}) and FHO 27 (Findeisen et al. ~\cite{find13}; Ibryamov et al. ~\cite{ibry19}). 
The "blueing effect" is observed during the minimums of brightness which is typical for the UXor variables.
Color-magnitude diagrams of V1180 Cas, like the ones of GM Cep and V1184 Tau, clearly show the "blueing" effect both in V-R and R-I colors (Mutafov et al. ~\cite{muta20};Mutafov et al. ~\cite{muta22}).
In our previous paper (Mutafov et al. ~\cite{muta19}) we concluded that the photometric properties of V1180 Cas can be explained by a superposition of highly variable accretion from the circumstellar disk onto the stellar surface and occultation from circumstellar clumps of dust, planetesimals or from other features of the circumstellar disk.
The new observational data confirms this conclusion but, additionaly, we observed significant change of the behaviour of the variability of the star, too - in the increase of its brightness, probably caused by increased accretion.
An attempt of finding a periodicity of light curves was made but without a satisfactory result.
The detailed explanation of this effect requires further observations.

\normalem
\begin{acknowledgements}
The authors thank the Director of Skinakas Observatory Prof. I. Papamastorakis and Prof. I. Papadakis for the award of telescope time. 
We thank Dr. Valentin D. Ivanov (ESO) for disscusions on the ideas for planetesimals.
This work was partly supported by the National RI Roadmap Project (contracts D01-383/18.12.2020 and D01-176/29.07.2022) of the Ministry of Education and Science of the Republic of Bulgaria.
This research has made use of the NASA's Astrophysics Data System Abstract Service, the SIMBAD database and the VizieR catalogue access tool, operated at CDS, Strasbourg, France. 

\end{acknowledgements}

\end{document}